\documentclass[aps,prl,reprint,superscriptaddress]{revtex4-2}
\usepackage{graphicx}
\usepackage{graphicx}% Include figure files
\usepackage{dcolumn}% Align table columns on decimal point
\usepackage{bm}% bold math
\usepackage[dvipsnames]{xcolor}
\usepackage{amsmath}

\begin{document}

%\preprint{}

\title{Pervasive beyond room-temperature ferromagnetism in a doped van der Waals magnet}



%\thanks{X.C. and Y.-T. S. contributed equally}

\author{Xiang Chen} 
\email{xiangchen@berbeley.edu}
\affiliation{Materials Sciences Division, Lawrence Berkeley National Lab, Berkeley, California 94720, USA}
\affiliation{Physics Department, University of California, Berkeley, California 94720, USA}

\author{Yu-Tsun Shao} 
\affiliation{School of Applied and Engineering Physics, Cornell University, Ithaca, New York 14853, USA}

\author{Rui Chen}
\affiliation{Department of Materials Science and Engineering, University of California, Berkeley, California 94720, USA}
\affiliation{Materials Sciences Division, Lawrence Berkeley National Lab, Berkeley, California 94720, USA}

\author{Sandhya Susarla}
\affiliation{Department of Materials Science and Engineering, University of California, Berkeley, California 94720, USA}
\affiliation{Materials Sciences Division, Lawrence Berkeley National Lab, Berkeley, California 94720, USA}

\author{Tom Hogan}
\affiliation{Quantum Design, Inc., San Diego, CA 92121, USA}

\author{Yu He}
\affiliation{Department of Applied Physics, Yale University, New Haven, Connecticut, 06511, USA}
\affiliation{Physics Department, University of California, Berkeley, California 94720, USA}
\affiliation{Materials Sciences Division, Lawrence Berkeley National Lab, Berkeley, California 94720, USA}

\author{Hongrui Zhang}
\affiliation{Department of Materials Science and Engineering, University of California, Berkeley, California 94720, USA}

\author{Siqi Wang}
\affiliation{NSF Nanoscale Science and Engineering Center (NSEC), 3112 Etcheverry Hall, University of California, Berkeley, California 94720, USA}

\author{Jie Yao}
\affiliation{Department of Materials Science and Engineering, University of California, Berkeley, California 94720, USA}
\affiliation{Materials Sciences Division, Lawrence Berkeley National Lab, Berkeley, California 94720, USA}

\author{Peter Ercius}
\affiliation{The Molecular Foundry, Lawrence Berkeley National Laboratory, Berkeley, CA 94720, USA}

\author{David A. Muller}
\affiliation{School of Applied and Engineering Physics, Cornell University, Ithaca, New York 14853, USA}
\affiliation{Kavli Institute at Cornell for Nanoscale Science, Cornell University, Ithaca, New York 14853, USA}

\author{Ramamoorthy Ramesh}
\affiliation{Department of Materials Science and Engineering, University of California, Berkeley, California 94720, USA}
\affiliation{Materials Sciences Division, Lawrence Berkeley National Lab, Berkeley, California 94720, USA}
\affiliation{Physics Department, University of California, Berkeley, California 94720, USA}

\author{Robert J. Birgeneau}
\email{robertjb@berkeley.edu}
\affiliation{Physics Department, University of California, Berkeley, California 94720, USA}
\affiliation{Materials Sciences Division, Lawrence Berkeley National Lab, Berkeley, California 94720, USA}
\affiliation{Department of Materials Science and Engineering, University of California, Berkeley, California 94720, USA}

%\footnote{* these authors contributed equally}
%\footnote{$^\dagger$ correspondence at: zxshen@stanford.edu, dunghai@berkeley.edu, mhashi@slac.stanford.edu}


\date{\today}

\begin{abstract}

The existence of long range magnetic order in low dimensional magnetic systems, such as the quasi-two-dimensional (2D) van der Waals (vdW) magnets, has attracted intensive studies of new physical phenomena. The vdW Fe$_N$GeTe$_2$ ($N$ = 3, 4, 5; FGT) family is exceptional owing to its vast tunability of magnetic properties. In particular, a ferromagnetic ordering temperature ($T_{\text{C}}$) above room temperature at $N$ = 5 (F5GT) is observed. Here, our study shows that, by nickel (Ni) substitution of iron (Fe) in F5GT, a record high $T_{\text{C}}$ = 478(6) K is achieved. Importantly, pervasive, beyond-room-temperature ferromagnetism exists in almost the entire doping range of the phase diagram of Ni-F5GT. We argue that this striking observation in Ni-F5GT can be possibly due to several contributing factors, including increased 3D magnetic couplings due to the structural alterations.

\end{abstract}

%\keywords{}

\maketitle


\section{Introduction}

Spontaneous symmetry breaking is forbidden at non-zero temperatures in isotropic spin systems with dimensions $d$ $\le$ 2 \cite{MerminWagner_1966_PRL, Hohenberg_1967_PRB}. Long range magnetic order in materials with reduced dimensionality can still be stabilized via both magnetic anisotropy and weak three-dimensional (3D) magnetic couplings \cite{Onsager_1944_PR, Lado_2017_2DM, Kim_2019_PRL}. In quasi-two-dimensional (quasi-2D) Heisenberg magnets, such as the van der Waals (vdW) bonded materials, the ordering process typically occurs as follows \cite{Als_Nielsen_1976_JPSSP, Chakravarty_1989_PRB, Birgeneau_1999_PRB}. At the highest temperature, the system is expected to exhibit 2D, classical isotropic magnetic correlations. As the temperature $T$ is lowered, the correlation length grows exponentially with $1/T$ and at sufficiently large length scales there is inevitably a crossover from 2D Heisenberg to 3D Ising or XY behavior followed by a phase transition to the 3D long range magnetic order. The details of the crossover depend on the strength of the 3D magnetic interactions and the symmetry and strength of the magnetic anisotropy. Van der Waals materials represent exciting realizations of these phenomena plus they contain the broad prospect of important technological applications \cite{KosterlitzThouless_1973, Park_2016_JPCM, Burch_2018_Nature, Gibertini_2019_NatNano, Gong_2019_Science, Wang_2020_AdP, Du_2021_NatRevPhy}.

Among the prominent bulk vdW materials for studying quasi-2D magnetism, such as Cr$_2$Ge$_2$Te$_6$ \cite{Carteaux_1995_JPCM, Gong_2017_Nature}, CrI$_3$ \cite{McGuire_2015_CM, Huang_2017_Nature}, Fe$_3$GeTe$_2$ (F3GT) \cite{Deiseroth_2006_EJIC_FGT, Chen_2013_JPSJ, Fei_2018_NatMat, May_2016_PRM_F3GT, Kim_2018_NatMat_Fe3GT, Zhang_2018_SciAdv} and CrTe$_2$ \cite{Freitas_2015_JPCM, Sun_2020_NanoRes}, the F3GT system is exceptional owing to the coupling between the electronic and magnetic degrees of freedom and its remarkable tunability. The bulk form of F3GT has a ferromagnetic (FM) transition temperature $T_{\text{C}}$ $\sim$ 230 K, which can be readily enhanced up to room temperature (RT), by either patterning the microstructure or applying ionic gating \cite{Deng_2018_Nature, Li_2018_NanoLett}. By intercalating more iron (Fe) into F3GT, $i.e.$, Fe$_N$GeTe$_2$ ($N$=4 for F4GT or 5 for F5GT) \cite{Seo_2020_SciAdv_F4GT, Stahl_2018_ZFAAC_F5GT, May_2019_ACSnano_F5GT}, the marked effects are the elevated $T_{\text{C}}$ up to RT and enhanced magnetization while only moderately increasing the magnetic moment size \cite{Seo_2020_SciAdv_F4GT, May_2019_ACSnano_F5GT, Zhang_2020_PRB_F5GT}. The F5GT compound is particularly interesting because of its advantageous characteristics for potential RT spintronic applications, including a $T_{\text{C}}$ above RT ($\sim$315 K), large magnetization ($\sim$700 kA/m), strong spin lattice coupling \cite{May_2019_ACSnano_F5GT, Zhang_2020_PRB_F5GT} and exotic magnetic textures \cite{Ly_2021_AFM, Gao_2020_AdvMat}.

To achieve practical applications of the quasi-2D magnetic materials, one common theme is to strengthen the FM and enhance the $T_{\text{C}}$ of the vdW magnets. Some proven avenues to dramatically raise the $T_{\text{C}}$ include gating \cite{Deng_2018_Nature, Verzhbitskiy_2020_NatNano}, applying pressure or strain \cite{Bhoi_2021_arxiv}, ion intercalation or carrier doping \cite{Weber_2019_NanoLett, Wang_2019_JACS_Cr2Ge2Te6}. By cobalt (Co) substitution of F5GT, $i.e.$ $({\text{Fe}}_{1-x}{\text{Co}}_x)_{5+\delta}{\text{GeTe}}_2$ (Co-F5GT), the magnetic ordering temperature is further increased to $\sim$ 360 K, along with the evolution of the magnetic ground state \cite{May_2020_PRM_FCGT, Tian_2020_APL_FCGT, Zhang_2021_FCGT, Zhang_2022_FCGT_sciadv}. More strikingly, at $x$ = 0.5 of Co-F5GT, a novel wurtzite-type polar magnetic metal was discovered, along with the N$\acute{\text{e}}$el-type skyrmion lattice at RT \cite{Zhang_2021_FCGT,Zhang_2022_FCGT_sciadv}. The aforementioned discoveries of the F5GT system highlight its immense tunability and capacity for applications in next-generation spintronics.

In this letter, we report pervasive, well-beyond-room-temperature FM in the F5GT vdW magnet with nickel (Ni) substitution, $i.e.$, $({\text{Fe}}_{1-x}{\text{Ni}}_x)_{5+\delta}{\text{GeTe}}_2$ (Ni-F5GT). Strikingly, a record high $T_{\text{C}}$ = 478(6) K is reached at a Ni doping level of $x = 0.36(2)$. In addition, the FM order persists robustly against Ni replacement until $x$ $\sim$ 0.86(1), beyond which only weak paramagnetism exists. Several factors might be relevant for the dramatic enhancement of the FM in Ni-F5GT.

\begin{figure}[t]
\centering
\includegraphics[width= 8.5 cm]{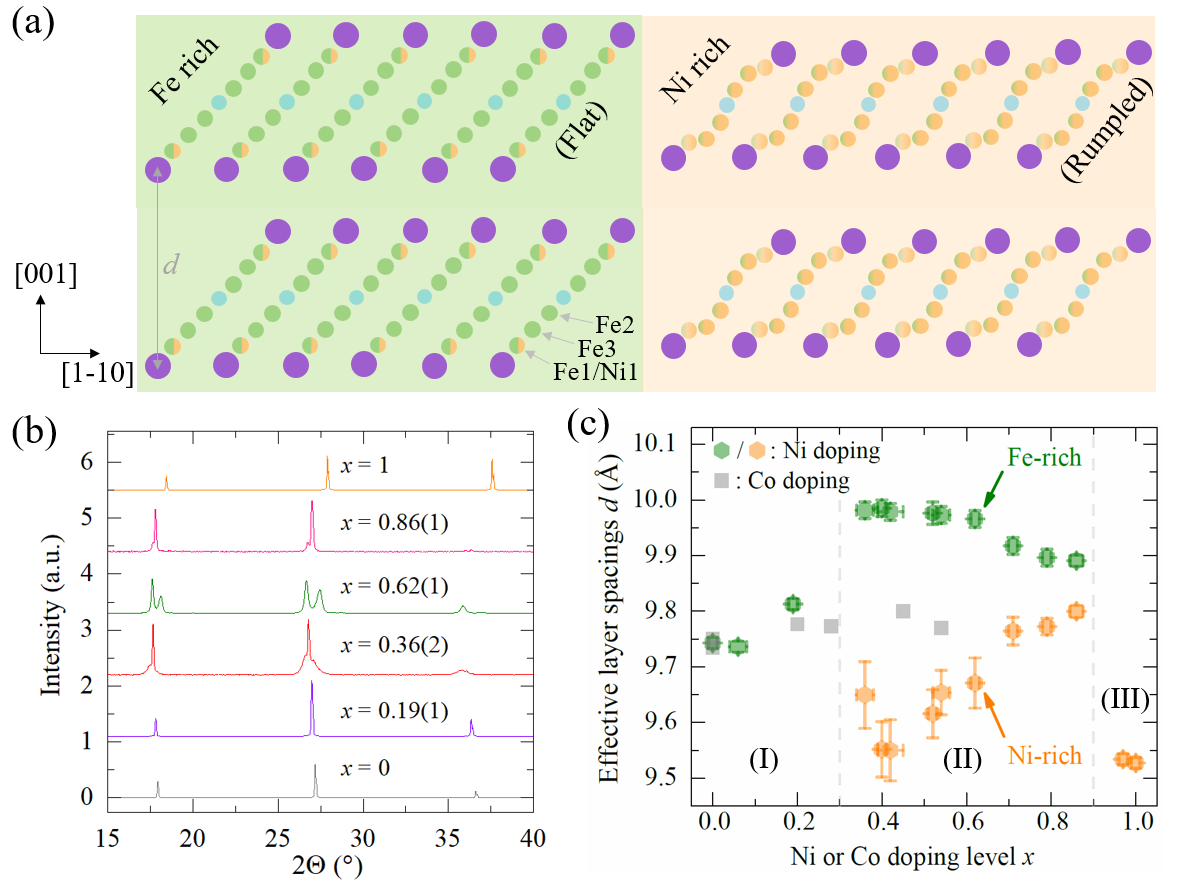}
\caption{(Color online) (a) Illustration of the AA-stacking order of both the Fe-rich and Ni-rich domains in Ni doped Fe$_5$GeTe$_2$ (Ni-F5GT) (Green symbols: Fe; Orange: Ni; Purple: Te; Blue: Ge) \cite{supp}. (b) The (0 0 L) type of peaks of Ni-F5GT at select Ni doping levels $x$. (c) Effective layer spacing $d$ as a function of Ni doping (colored symbols) or Co doping (gray symbols, \cite{Zhang_2021_FCGT}) in Fe$_{5+{\delta}}$GeTe$_2$. Vertical dashed lines indicate different regions in Ni-F5GT.}
\label{fig:Fig1_alpha}
\end{figure}

The Ni-F5GT single crystals were grown by the chemical vapor transfer method \cite{May_2019_ACSnano_F5GT, Zhang_2020_PRB_F5GT, supp}. The exact Ni doping level $x$ and total cation count per formula unit ($f.u.$) $5+\delta$ were verified by energy-dispersive X-ray spectroscopy (EDX/EDS) (Fig. S1). The lattice and atomic structure of Ni-F5GT were investigated by a combination of techniques, including powder and/or single crystal x-ray diffraction (XRD) and high-angle annular dark-field scanning transmission electron microscopy (HAADF-STEM). Magnetization measurements were performed via a commercial Quantum Design MPMS3.

\begin{figure*}[t]
\centering
\includegraphics*[width=16.2 cm]{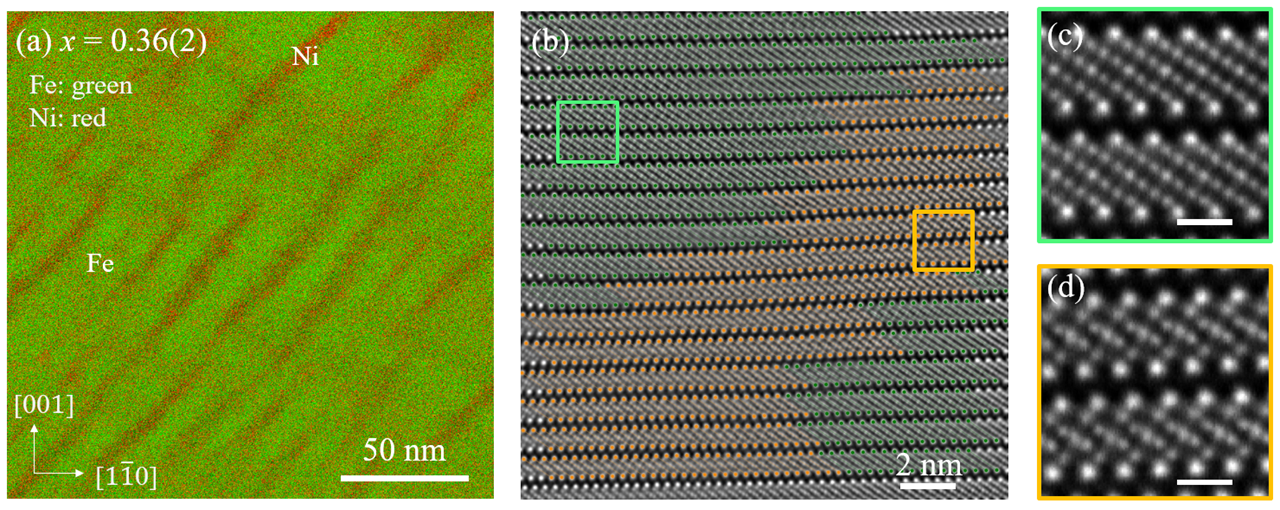}
\caption{(Color online) Nanoscale phase separation of Ni-F5GT at $x$ = 0.36(2). (a) Energy-dispersive X-ray spectroscopy map showing both the Fe-rich (green) and Ni-rich (red) regions. (b)-(d) Atomic resolution HAADF-STEM image demonstrating the two types of domains. The enlarged (c) Fe-rich and (d) Ni-rich domains show flat and rumpled atomic planes, respectively. The Te-Te planes in (b) are color-coded based on the number and the rumpling of the atomic layers. Scale bars in (c)-(d), 5 \AA. }
\label{fig:Fig2_alpha}
\end{figure*}

The unit cell of F5GT is composed of three identical layers (stacks) with the rhombohedral layer stacking (space group R$\bar{3}$m), labelled as ABC-stacking \cite{May_2019_ACSnano_F5GT}. Upon introducing Ni into F5GT, such as at $x$ = 0.19(1), the crystal structure undergoes a transition from ABC-stacking to AA-stacking (space group P$\bar{3}$m1, illustrated in Fig. 1(a)), as confirmed by single crystal XRD (Fig. S2). The AA-stacking order is also verified by HAADF-STEM images at other doping levels (Fig. 2 and Fig. S5). When viewed along the [1 1 0] direction, the identical layers stack exactly on top of each other along the $c$ direction and are separated by a vdW spatial gap. Within each layer (stack), the center germanium (Ge, blue symbols) atom is surrounded by three different sites of iron (Fe1, Fe2 and Fe3, green) atoms which are further protected by the outer Te (purple) atoms, $i.e.$, forming a Te-Fe1-Fe3-Fe2-Ge-Fe2-Fe3-Fe1-Te plane-like (labelled as TGT-$plane$) structure.

 With increasing Ni doping in Ni-F5GT, the excess cation count $\delta$ gradually increases from $\delta \sim$ 0 at $x$ = 0 to $\delta$ $\sim$ 0.5 at $x$ $\sim$ 0.3 (labelled as region (I) with 0 $\le$ $x$ $\le$ 0.3), and saturates when $x$ $>$ 0.3 (Fig. S1). Interestingly, in the intermediate doping range 0.36(2) $\le$ $x$ $\le$ 0.86(1) (region (II)), although the sample still maintains the AA-stacking, two types of domains coexist, as evidenced from the splitting of the (0 0 $L$) peaks from the XRD measurements (Fig. 1(b)), the direct atomic visualization from the HAADF-STEM images (Fig. 2 and Fig. S5) and the STEM-EDX maps (Fig. 2 and Fig. S4). As an example at $x$ = 0.36(2) (Fig. 2 and illustrated in Fig. 1(a)), the two types of domains correspond to Fe-rich domains (Fig. 2(c)) and Ni-rich domains (Fig. 2(d)), respectively. In the Fe-rich domains, the TGT-$planes$ are almost flat and more extended in space; while the TGT-$planes$ are rumpled in the Ni-rich domains. Because of the contrasting local atomic arrangements of the Fe and Ni atoms within the TGT-$planes$, the effective layer spacing $d$ of either the Fe-rich or Ni-rich domain shows a strong deviation from the value at $x$ = 0 (Fig. 1(c)). This strong alteration may have a dramatic impact on the electronic and/or magnetic properties of Ni-F5GT.

While the Ni-F5GT samples are still metallic (Fig. S6), the magnetic properties are strongly influenced by Ni substitution (Fig. 3 and Fig. S7). With moderate Ni replacement, the $T_{C}$ of Ni-F5GT is already considerably enhanced (Figs. 3(a)-3(b)). For instance, at $x$ = 0.19(1), the experimentally determined $T_{C}$ = 395(6) K already approaches 400 K (Fig. 3(a)). Meanwhile, the Ising spin moment switches to the in-plane direction and seems to remain so until $x$ $\sim$ 0.86(1) (Fig. 3(e) and Fig. S7). Upon further increasing the Ni content, a record high $T_{C}$ $=$ 478(6) K is achieved at $x$ = 0.36(2), together with the maximum in-plane ($H_{C}^{ab}$ $\approx$ 500 Oe) and out-of-plane ($H_{C}^{c}$ $\approx$ 1600 Oe) coercive fields (Figs. 3(c), 3(f) and Fig. S7). After reaching the maximum $T_{C}$ in Ni-F5GT, the FM order is only gradually weakened by further Ni doping. Strikingly, even at $x$ = 0.86(1), the as-grown sample of Ni-F5GT maintains an above-room-temperature $T_{C}$ $\approx$ 380 K, which is only lowered and stabilized at 220 K or 150 K, depending on how the sample is thermally cycled above its original $T_{C}$ (Fig. S8). Only beyond this point of further Ni replacement (0.86(1) $<$ $x$ $\le$ 1, region (III)), the FM order of Ni-F5GT is completely suppressed, rendering the weak paramagnetism, accompanied by the adoption of a different, layered tetragonal structure (space group I4/mmm) as Ni$_{5.5}$GeTe$_2$ \cite{Deiseroth_2007_ZfK_Ni5GT}.

The saturation magnetic moment per $f.u.$ of Ni-F5GT at 2 K decreases approximately linearly with increasing Ni content (Figs. 3(c),(e) and Fig. S7), from $\sim$10 $\mu_B$/$f.u.$ at $x$ = 0 to nearly zero (weakly paramagnetic) at $x$ = 1. This implies that the Ni dopants are not magnetic and only dilute the FM in Ni-F5GT. Surprisingly, the magnetic ordering temperature $T_{C}$ does not follow this expected trend (Fig. 4). Instead, pervasive, above-room-temperature FM exists almost over the entire range of the phase diagram (0 $\le$ $x$ $\le$ 0.79(1)). Particularly, in region (I) of the phase diagram of Ni-F5GT, the $T_{C}$ shows a strong, positive deviation from the Vegard's Law behavior ($i.e.$, a dilution of the magnetic component), thus suggesting additional factors might be present for the unusual evolution of $T_{C}$. Our study demonstrates the remarkable Ni enhanced FM in Ni-F5GT, as summarized in Fig. 4, where three different regions are categorized based on the structural and magnetic characterization.

\begin{figure*}[t]
\centering
\includegraphics*[width=17.2 cm]{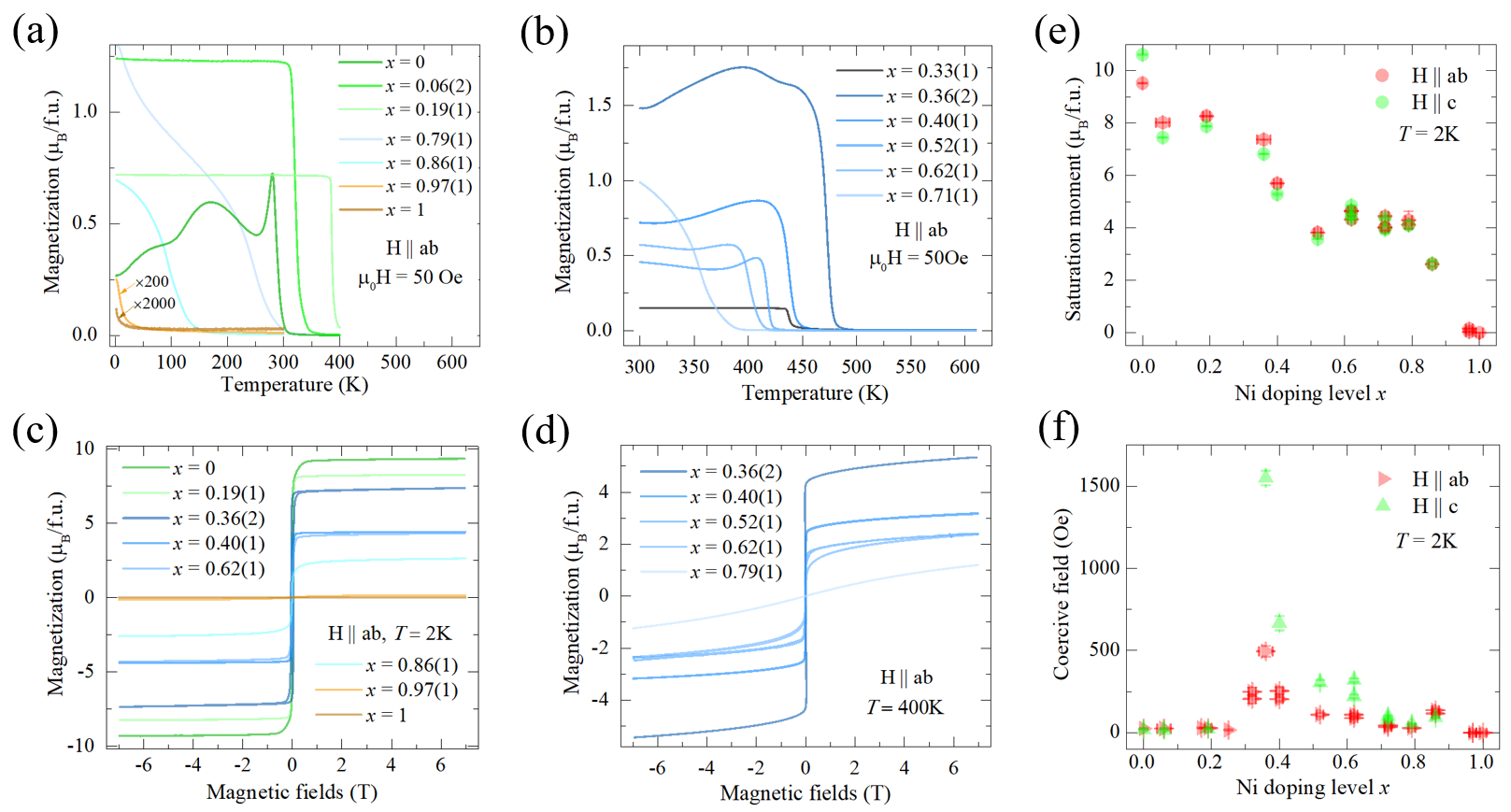}
\caption{(Color online) Magnetization data of Ni-F5GT at select Ni doping level $x$. (a)-(b) Temperature dependent magnetization of Ni-F5GT (external in-plane magnetic field is 50 Oe). For clarity, the magnetization data at $x$ = 0.97(1) and $x$ = 1 are multiplied by 200 and 2000, respectively. (c)-(d) In-plane isothermal magnetization at $T$ = 2 K (c) and $T$ = 400 K (d), respectively. (e) Saturation moment per formula unit at 2 K under the magnetic field of 7 T. (f) Coercive fields extracted from the isothermal magnetization at $T$ = 2 K. In (e)-(f), the magnetic field is applied along the $ab$-plane (red symbols) or $c$-direction (green symbols).}
\label{fig:Fig3_alpha}
\end{figure*}

% Now explain the magnetism

The unique feature of Ni-F5GT is the immense impact on the lattice and magnetism by Ni substitution, including the radical change of the layer spacing $d$ (Fig. 1(c)) and the local atomic arrangements (Fig. 1(a), Fig. 2, Figs. S2-S5), as compared to F5GT or Co-F5GT \cite{May_2019_ACSnano_F5GT, Zhang_2021_FCGT}. It is evident that the change of the layer spacing $d$ in Ni-F5GT is significantly more pronounced than its evolution in Co-F5GT. Importantly, the layer spacing $d$ of the Fe-rich domains (green symbols in Fig. 1(c)) closely tracks the evolution of $T_{\text{C}}$ in Ni-F5GT at 0 $\le$ $x$ $\le$ 0.86(1) (Fig. 4). This is supported by the experimental observation that the Fe-rich domains maintain the AA-stacking order with the straight-and-flat TGT-$planes$ over a broad Ni doping range 0 $<$ $x$ $\le$ 0.86(1) and thus may be mainly responsible for the robust FM in Ni-F5GT. The Ni dopants are also indispensable for donating electrons to the system and forming the Ni-rich domains in region (II), which help preserve the Fe-rich domains. This phase separation in region (II) naturally explains the domain pinning enhanced coercive fields (Fig. 3(f)). Meanwhile, the preserved Fe-rich domains maintain the robust $T_{C}$ of Ni-F5GT in this region.

Our work on Ni-F5GT reveals a complicated yet intriguing system, in which the lattice, electronic and magnetic degrees of freedom are closely intertwined. To understand the enhancement of FM in Ni-F5GT, especially the region (I) (Fig. 4), it is necessary to consider the possible effects of both electron doping and structural alteration by Ni replacement. Carrier doping is often seen to be detrimental to the correlation induced long range magnetic order, as recognized in the unconventional superconductors, such as the cuprates and iron pnictides \cite{Lee_2006_RMP, Dai_2015_RMP}. For the itinerant FM, charge doping can, in some cases, help meet the Stoner criterion and actually promote magnetic order \cite{Huang_2021_PRB}. The related F3GT system, which may also apply to F5GT, has both itinerant and localized spin moment contributions to the magnetism \cite{Chen_2013_JPSJ, May_2016_PRM_F3GT, Deng_2018_Nature, Kim_2018_NatMat_Fe3GT, Zhang_2018_SciAdv, Calder_2019_PRB, Xu_2020_PRB_Fe3GT_ARPES}. Hence, the ordering temperature of F3GT can be elevated by electrostatic gating of its thin layer form \cite{Deng_2018_Nature, Verzhbitskiy_2020_NatNano}. In Ni-F5GT, the electrons provided by the Ni dopants may greatly influence the electronic band structure and the density of states (DOS) near the Fermi level ($E_{\text{F}}$). Alternatively, the magnetic exchange coupling $J_{i,j}$ between spins $\boldsymbol{S}_i$ and $\boldsymbol{S}_j$ on sites $i$ and $j$, might also be altered via the Ruderman–Kittel–Kasuya–Yosida (RKKY) exchange since the electronic structure is likely altered substantially by Ni-doping \cite{Deng_2018_Nature, Zhang_2018_SciAdv, Seo_2021_NatComm_CoFe4GT, Zhao_2021_NanoLett}. All together, the itinerant FM might also be enhanced by electron doping in Ni-F5GT.

Now focusing on the localized spin moments and considering a simple Heisenberg model with a weak magnetic anisotropy, in which the magnetic contributions from the itinerant FM can also be effectively mapped into the Hamiltonian \cite{Prange_1979_PRB, Deng_2018_Nature}:

\begin{equation}
 H =  \sum_{\substack{i < j}} J_{i,j} \boldsymbol{S}_i \cdot \boldsymbol{S}_j 
     - \sum_{\substack{i}} A (S_i^z)^2
\label{eq:eqn1}
\end{equation}

Here, $A$ is the single-ion anisotropy ($A$ $>$ 0 for Ising spin moment). Since $A$ is small and close to zero in Ni-F5GT \cite{Zhang_2020_PRB_F5GT}, a mean-field treatment results in the magnetic transition temperature \cite{Deng_2018_Nature, Gibertini_2019_NatNano}:

\begin{equation}
  T_C = \frac{ S(S+1)}{3k_{B}} ( {z_{nn}} J_{nn}  + ... )
\label{eq:eqn2}
\end{equation}

where $z_{nn}$ is the coordination number of the nearest neighbouring sites, $S$ is the magnetic spin quantum number and $J_{nn}$ the nearest-neighbour exchange coupling. On the mean-field level, a larger $z_{nn}$ or $J_{nn}$ promises a larger $T_{C}$. Perhaps, this is why the ordering temperature in Fe$_N$GeTe$_2$ is quickly increased from $T_{C}$ $\sim$ 230 K at $N$ = 3 to $T_{C}$ $\sim$ 317 K at $N$ = 5 \cite{Deiseroth_2006_EJIC_FGT, Chen_2013_JPSJ, Seo_2020_SciAdv_F4GT, May_2019_ACSnano_F5GT, Zhang_2020_PRB_F5GT}. In Ni-F5GT, since the average magnetic moment per Fe is almost unchanged and the in-plane FM indicates a small yet negative $A$ (Fig. 3(e) and Fig. S7), hence neither $S$ nor $A$ is responsible for the enhancement of the FM. However, the structural alterations may affect both $z_{nn}$ and $J_{i,j}$, and therefore are strong candidates for explaining the further enhanced $T_{C}$.

Firstly, a small increment of $\delta$ is observed for lightly Ni doped F5GT (region (I) in Fig.~4 and Fig. S1). One direct consequence is the larger site occupancy of the Fe1 site, which is up to $\sim$75$\%$ at $x$ $\sim$ 0.3 as compared to a maximum of $\sim$50$\%$ at $x$ = 0. Considering that the Fe1-Fe3 bond length is the shortest among all of the direct Fe-Fe bonds (Fig. S3), the Fe1 site might be critical for determining the $T_{C}$ in Ni-F5GT. Therefore, a larger $\delta$, which implies a greater $z_{nn}$ of the Fe1 site, might promote a higher $T_{C}$ (Eqn. (2)). Secondly, with more Ni substitution in region (I), the explicit effect is that the TGT-$planes$ of the Fe-rich domains are becoming more flat and extended in space (Fig. 2(c), Fig. S3, Figs. S5(c),(e)). Microscopically, other than the small spatial re-arrangements of the Fe2 and Fe3 sites, it is the Fe1 site that is becoming increasingly distant from the Fe3 site along the $c$ direction while keeping the Fe-Fe bond lengths nearly unchanged (Fig. S3). This explains the enlargement of the layer spacing $d$ (Fig. 1(c)), which positively correlates with the $T_{C}$ in Ni-F5GT. Understandably, the intralayer and interlayer exchange coupling, especially these related to the Fe1 site, might be effectively strengthened, which render the 3D magnetic interactions stronger and eventually lead to the enhancement of $T_{C}$ in Ni-F5GT.

\begin{figure}[t]
\centering
\includegraphics*[width= 7.2 cm]{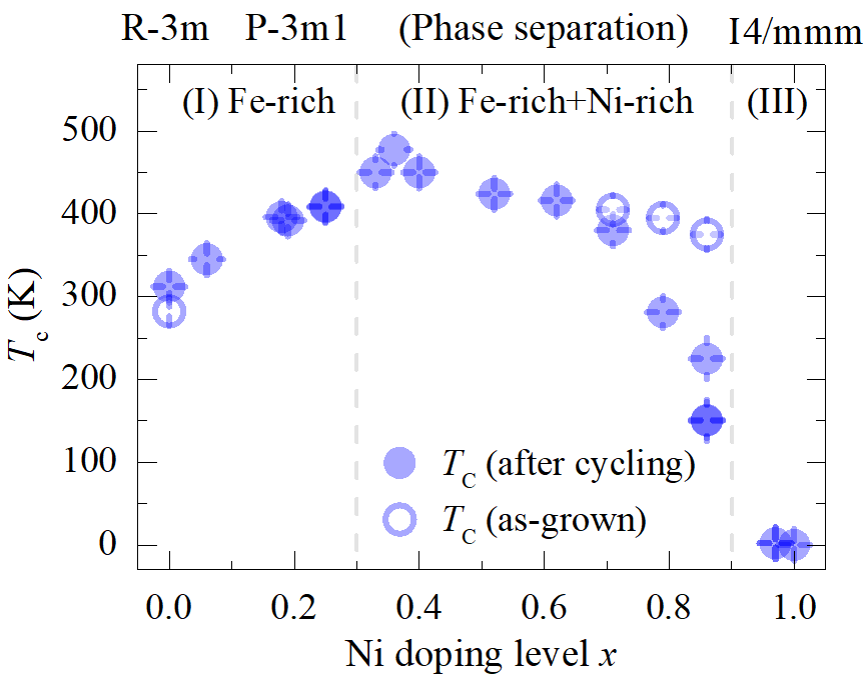}
\caption{(Color online) Ni doping level $x$ dependent $T_{C}$ in Ni-F5GT. Empty symbols: $T_{C}$ for as-grown samples; solid symbols: $T_{C}$ after thermal cycling. Vertical dashed lines indicate different regions in Ni-F5GT.}
\label{fig:Fig4_alpha}
\end{figure}

In summary, our work reveals a new arena for studying the vdW magnetic metals with strong room temperature ferromagnetism. A record high ferromagnetic order with a $T_{C}$ $=$ 478(6) K is realized in Ni-F5GT at $x$ = 0.36(2). Albeit with the non-magnetic dilution, several factors are speculated to assist the pervasive, well-above-room-temperature FM in Ni-F5GT. Candidate contributors include the increased site occupancy, structural modifications altered magnetic exchange couplings and the electron doping effect. Clearly, as stated, these ideas are purely speculative and require much more thorough investigation. Although further research is needed to understand fully the mechanism of the enhanced magnetism, our study highlights that the Ni-F5GT system is an extremely rare example of strongly enhanced ferromagnetism, in spite of the detrimental factors such as the non-magnetic dilution and electron doping effects introduced by Ni dopants. In addition, Ni-F5GT offers unique or alternative avenues towards enhanced coercivity, varying length scale of phase separation, thermal cycling influenced $T_{C}$ and potential relevance to skyrmionics in F5GT and other related vdW magnets \cite{May_2020_PRM_FCGT, Tian_2020_APL_FCGT, Zhang_2021_FCGT, Zhang_2022_FCGT_sciadv}.

\bigbreak

X.C. wishes to thank Nicholas S. Settineri and Weiwei Xie for some single crystal XRD measurement and acknowledge the Applications Group at Quantum Design for their contribution of high temperature (300-600 K) magnetization measurements to this work. Work at Lawrence Berkeley National Laboratory was funded by the U.S. Department of Energy, Office of Science, Office of Basic Energy Sciences, Materials Sciences and Engineering Division under Contract No. DE-AC02-05-CH11231 within the Quantum Materials Program (KC2202). Y.T.S., H.Z. and D.A.M acknowledge financial support from the Department of Defense, Air Force Office of Scientific Research under award FA9550-18-1-0480. R.C. and J.Y. acknowledge the support by Intel Corporation under an award titled Valleytronics center. The electron microscopy studies were performed at the Cornell Center for Materials Research, a National Science Foundation (NSF) Materials Research Science and Engineering Centers program (DMR 1719875). The microscopy work at Cornell was supported by the NSF PARADIM DMR-2039380, with additional support from Cornell University, the Weill Institute and the Kavli Institute at Cornell. S.S. acknowledges the help from Dr. Rohan Dhall and is supported by the Quantum Materials Program under the Basic Energy Sciences, Department of Energy. The microscopy work was performed at Molecular Foundry that is supported by the Office of Science, Office of Basic Energy Sciences, of the U.S. Department of Energy under Contract No. DE-AC02-05CH11231. The devices for transport measurements were fabricated in the UC Berkeley Marvell Nanofabrication Laboratory.

\bibliography{FeNi_FGT}

\end{document}